# Towards Semantic Big Graph Analytics for Cross-Domain Knowledge Discovery

Feichen Shen

School of Computing and Engineering,
University of Missouri- Kansas City

**Abstract**. In recent years, the size of big linked data has grown rapidly and this number is still rising. Big linked data and knowledge bases come from different domains such as life sciences, publications, media, social web, and so on. However, with the rapid increasing of data, it is very challenging for people to acquire a comprehensive collection of cross domain knowledge to meet their needs. Under this circumstance, it is extremely difficult for people without expertise to extract knowledge from various domains. Therefore, nowadays human limited knowledge can't feed the high requirement for discovering large amount of cross domain knowledge. In this research, we present a big graph analytics framework aims at addressing this issue by providing semantic methods to facilitate the management of big graph data from close domains in order to discover cross domain knowledge in a more accurate and efficient way.

## 1. Problem Statement

Today, the main challenge we are facing in knowledge discovery research is the big data problem associated with large, complex, and dynamic variation of format. Increasingly, we are also seeing the emergence of cross domain among different datasets. In this drive, the large amounts of data have been specified and shared via machine-readable formats, such as a Resource Description Framework (RDF) [1] and Ontology Web Language (OWL) [2].

However, with increasing research in big data, more and more datasets from different domains have been added to the existing linked open data (e.g., Bio2RDF [3]), which makes the highly complex relationships and condensed interlinks among the large number of these knowledge bases. To some extent, the speed of data growing in terms of multiple domains is much faster than that of the large amount of knowledge people can acquire and consume in their daily lives. In other words, since cross domain datasets are physically grouped instead of semantically clustered, it is extremely difficult for people without expertise to extract knowledge from various domains nowadays. Therefore, there is a big gap between human limited knowledge and the large amount of cross domain knowledge that can be discovered from this huge amount of data.

For example, in life science research, a researcher expertise on disease A wants to find the corresponding ensemble genome G for the gene symbol and marker that re-lated to disease A as well as the drug D for disease A. In this use case, A, G and D represent three different domains, such as Disease Ontology [4], Gene Ontology [5], and DrugBank [6]. More knowledge bases and ontology exist in the life science domain, including the Human Phenotype Ontology [7], the Mammalian Phenotype Ontology [8, 9], the Pharmacogenetics Knowledge Base (PharmGKB) [10], the Onine Mendeian Inheritance in Man (OMIM) [11], the Database of Chromosomal Imbalance and Phenotypc in Humans Using Ensembl Resources (DECIPHER) [12], the Orphanet [13] and so on. With tremendous heterogeneous knowledge graphs, it is very difficult for researchers who only know disease domain to make a query to discover knowledge across different domains in a comprehensive way.

**The problem** we intended to address is to fill the gap between human limited knowledge with the huge amount of cross-domain knowledge by providing semantic methods to facilitate the partition and management of big cross-domain data in a more accurate and efficient way. We assume that each output group of our approach contains knowledge sources that have most closeness relationship. We can design cross-domain query based on each group to dig the comprehensive relationship among knowledge sources in it to fully discover cross-domain knowledge.

## 2. Relevancy

The problem is directly relevant to *graph partition* and *graph analytics* with an emphasis on semantic knowledge discovery [14-19] and graph based reasoning [20-22]. In *graph partition*, domain knowledge in RDF format can be partitioned not only physically but also semantically. Fuzzy partition is proposed in this research, which means that overlap of nodes is permitted since a same node can be reused for multiple domains. In *graph analytics,* how to group different domain knowledge together is based on graph measurement and analysis (e.g., distance, similarity). A semantic based similarity measurement is proposed in this research and unsupervised clustering approach is used to form different groups of close domain knowledge.

## 3. Related Work

In this section, we present previous work in graph schema partition and analytics.

### 3.1 Graph schema partition

SEDGE [23] provides a complementary partition approach to eliminate cross domain edges to facitilty query performance. SEDGE also proposed an on-demand parititon to handle unbalanced query workload. Unfortunately, the partition is mainly based on physical relationships rather than semantic relationships.

Mizzan [24] made improvements based on Pregel [25] that is built on Bulk Synchronous Parallel (BSP) [26] programming model. Mizzan focuses on dynamically efficient load balancing in terms of computation and communication among all worker nodes. It achieved load balancing by using fine-grained vertex migration in a distributed manner. However, Mizzan designed a vertex centric model, mainly focused on the size balance load of the graph rather than the semantic data migration.

Goffish [27] is a distributed sub-graph centric approach using a connected component approach to make abstraction for a large scale graph in order to do efficient graph analytics. This model combined the advantages of both a vertex centric approach and shared-memory algorithms. However, Goffish did not mention semantic cross domain issues. In Similarity-Driven Semantic Role Induction via Graph Partitioning [28], authors proposed a vertex centric unsupervised method for semantic role induction. But they did not mention cross domain issues. In this sense, our approach provides a better solution to partition large graphs by grouping semantically related content together in sub-graphs.

### 3.2 Graph Analytics

We categorize research in this area into two parts: query generation and query processing. For query generation, SP$^2$Bench [29] proposed a query design system focusing on generating queries with combination of different operations. But this query generation was not designed from a semantic perspective for cross domain. LUBM [30] and BSBM [31] generated benchmarks on university and ecommerce respectively, but neither benchmark was based on more than one domain. FedBench [32] provided a benchmark suite for federated queries on semantic data which can cover semantic multiple domain data use cases. However, query benchmarks were manually generated by authors. Our approach provides a way to automatically help people find the semantic relationship without acquiring knowledge explicitly. MedTQ provided a dynamic topic query generation approach over medical ontologies [33]. BmQGen [34, 35] supported query generation for a specific biomedical detecting task on postsurgical complications [36]. Zhu et al. investigated the usage of the PharmGKB knowledge base for searching and inferring repositioning breast cancer drugs [37].

For query processing, H2RDF+ [38] provided a scalable distributed RDF store to facilitate query processing performance. Trinity [39] presented a RDF management framework over a distributed in-memory key-value store. There are still some other indexing approaches for speeding query processing like RDF-3X [40] and gStore [41], which can reduce graph searching space and avoid looking up unnecessary blocks. However, neither of them works on a cross domain and semantic perspective. Indexing technology in our approach provides semantic meaning for graphs.

### 4. Research Questions

The main goal of my doctoral research is to build a big graph analytics framework for cross-domain knowledge discovery. In order to achieve this goal, the research will give answers to the following research questions:
  (i)     How to provide an effective and accurate semantic partitioning scheme to discover knowledge?
  (ii)    For RDF data, how can we determine the closeness relationships, e.g., based on predicate-predicate or subject/object-subject/object relationships?
  (iii)   How to build a scalable framework to process and manage large data?

## 5. Hypotheses

The main hypotheses related to my research are:
- Any normailized RDF data from heterogeneous resources can be the input of our framework
- Edge centric similarity results represent similarity of knowledge domain.
- Different knowledge sources belong to one of the clustering types must have a closer relationsihp than the ones not in the same cluster/group.
- Data is big and increases infinitely, there is no way to handle that huge amounts of data in a standalone machine.

## 6. Approach

We found out that predicates play an important role as hubs to share information and connect entities among heterogeneous data. Therefore, we hypothesize that graphs can be fuzzy clustered based on a predicate sharing and distance measurement, and data in the same clustered group have closer relationships than when separate.

In this drive, we defined an edge centric neighboring algorithm to determine the closeness of any two predicates. The formal definition is shown in Definition 1.

**Definition 1:** Given a directed graph G (V, E), vertices $V_s$, $V_p$, $V_o$ denote subject, predicate, and object nodes in the RDF schema graph, respectively. Let d $(V_{pi}, V_{pj})$ represent the shortest path between $V_{pi}$ and $V_{pj}$, r $(V_{pi}, V_{pj})$ determines the reachability between $V_{pi}$ and $V_{pj}$, n $(V_{pi}, V_{pj})$ indicates the neighbors' closest level between $V_{pi}$ and $V_{pj}$:

$$n(V_{pi}, V_{pj}) = \begin{cases} 1, & if\ d(V_{pi}, V_{pj}) = 1 \\ L, & if\ d(V_{pi}, V_{pj}) = L\ (L > 1)\ AND\ r(V_{pi}, V_{pj}) = true \end{cases}$$

After assigning levels to different pair of predicates, we utilize the clustering approach for discovering the predicate association patterns from the ontologies. The similarity based distance measurement for the clustering algorithm varies based on different neighboring levels for each pair of predicates. Basically, we give higher weights to closer predicates and lower weights to further predicates. That is, for any two predicates in level 1 relationship, we assign a probability based similarity score to them. The definition is given in Definition 2.

**Definition 2:** Given $V_{pi}$ and $V_{pj}$ in a directed RDF schema Graph G (V, E). Let A and B denote the number of entity (subject or object) directed connected to $V_{pi}$ and $V_{pj}$ regardless of the direction respectively. ps $(V_{pi}, V_{pj})$ indicates the probability based similarity between $V_{pi}$ and $V_{pj}$:

$$ps(V_{pi}, V_{pj}) = \frac{A \cap B}{A} * \frac{A \cap B}{B}$$

Moreover, assigning the same similarity score for different level of neighboring relationship is unfair. In this case, we define a dynamic weight for probability based similarity in different levels in Definition 3.

**Definition 3:** Denote n ($V_{pi}$, $V_{pj}$) = m (m>1), and n ($V_{pi}$, $V_{pk}$) = p (1≤p<m), n ($V_{pk}$, $V_{pj}$) = q (1≤q<m)

$$ps(V_{pi}, V_{pj}) = \begin{cases} ps(V_{pi}, V_{pk}) * ps(V_{pk}, V_{pj}), & if\ m = 2 \\ Max(ps(V_{pi}, V_{pk}), ps(V_{pk}, V_{pj})) & if\ m > 2 \end{cases}$$

After we get the similarity score for all pairs of predicates, we use formula in Definition 4 to generate a distance matrix for clustering.

**Definition 4:** Given distance matrix CM and total number of predicate n. Denote $ps_{ij}$ as the probability-based similarity score between predicates $p_i$ and $p_j$ based on different levels, so that:

$$CM[p_i, p_j] = \begin{cases} PS_{ij}, & if\ p_i \neq p_j, 0 \leq i \leq n, 0 \leq j \leq n \\ 1 & if\ p_i = p_j, 0 \leq i \leq n, 0 \leq j \leq n \end{cases}$$

To discover the correlation between predicates, we used an innovative Hierarchical Fuzzy C-means (HFCM) clustering algorithm. We created HFCM algorithm and made a functional extension based on a Fuzzy C-means clustering algorithm. In general, we set a machine capacity threshold to denote a certain number of triplets that each machine can hold. In addition, we continue applying an HFCM algorithm on each cluster until the number of triplets for each cluster is less than or equal to the threshold or no further change of numbers of elements for each cluster can be made. Moreover, to get the optimal number of clusters, we use Silhouette Width to evaluate different results and choose the one with the biggest score.

## 7. Preliminary Results

The preliminary evaluation is conducted in terms of the validation of clustering result and justified query benchmark generation. We used eight ontologies from Bio2RDF release 3 to evaluate our system. Detailed is given in Table 1. In addition, we eliminated some RDF built-in predicates and types for getting the best clustering result. The total number of predicate for HFCM reduced from 1099 to 1064.

Table 1: Bio2RDF Datasets

| Ontology | Triples# | # of Unique Triple Schema | # of Unique Predicates |
|---|---|---|---|
| Biomodels | 2 million | 478 | 18 |
| BioPortal [42] | 20 million | 8259 | 867 |
| DrugBank | 3 million | 737 | 68 |
| GOA | 97 million | 75 | 17 |
| HGNC [43] | 3 million | 57 | 18 |
| MGI [44] | 8 million | 95 | 19 |

| OMIM | 8 million | 261 | 39 |
| Pharmgkb | 270 million | 396 | 53 |
| **Total** | **411 million** | **10258** | **1099** |

Firstly, we conduct heuristic comparison experiments among different clustering algorithms to choose an appropriate approach to get the best number of clusters. Figure 1 shows an edge centric clustering results for probability based similarity in terms of running five different clustering algorithms (Fuzzy C-means, K-means, Clara, Pam and Hierarchical Clustering) on the input data. We used silhouette width as the validation method to do evaluation. As a result, Clara, Pam and hierarchical algorithms give a relative stable silhouette width for each number of clusters. This means there is no optimal cluster number for these cases. Both Fuzzy C-means and K-means give the highest silhouette width 0.95 when the cluster number is 5. However, we choose Fuzzy C-means because it gives additional soft partition information for each cluster, which could be useful for distributed query processing.

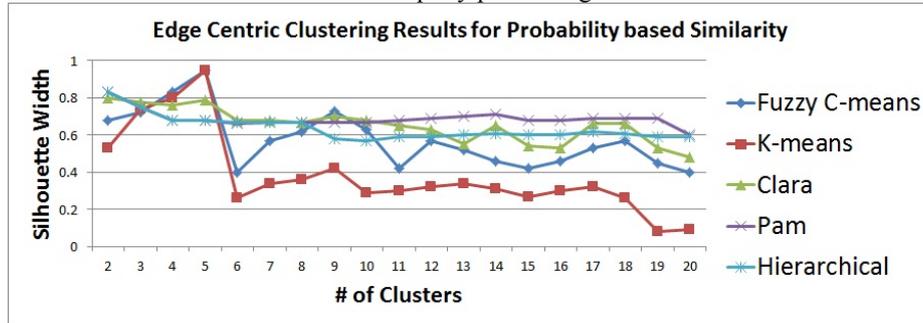

**Fig. 1.** Edge Centric Clustering Results for Probability based Similarity

After finalizing Fuzzy C-means with a probability based similarity score as the measurement approach, we apply a hierarchical Fuzzy C-means (HFCM) algorithm to the dataset trying to partition each cluster into an indivisible unit. For each level of processing, we still used silhouette width to determine the proper number of clusters. As Figure 2 shows, in first level running of HFCM, we get five clusters. Then in second level running of HFCM, only cluster 2 (C2) and cluster 3 (C3) can be further clustered. According to silhouette width, we get the best number of cluster 2 for C2 and 4 for C3 respectively. As a result, after level 2 running, we get 9 different clusters. Because the number of cluster does not change as we run the $3^{rd}$ level HFCM, the algorithm stops here. The solutions from the algorithm yield the number of clusters as 9.

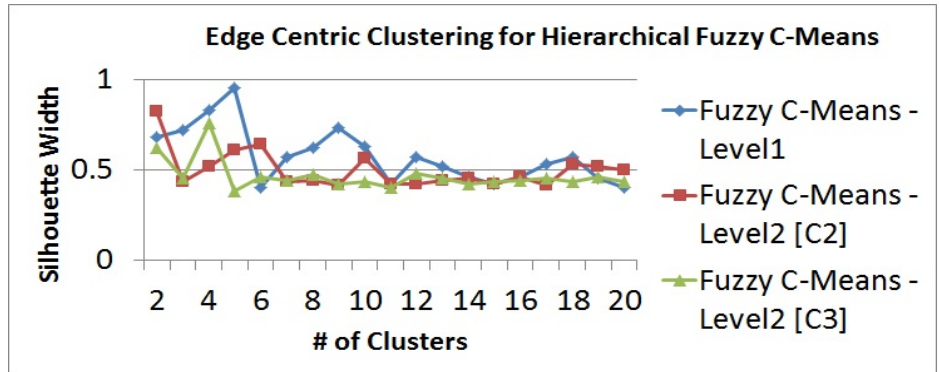

**Fig. 2.** Edge Centric Clustering for Hierarchical Fuzzy C-Means

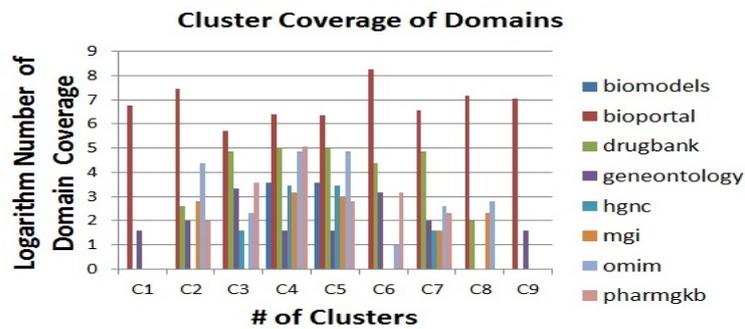

**Fig. 3.** Cluster Coverage of Domain Knowledge

We conducted a domain coverage evaluation for each cluster. Figure 3 shows such results with histogram. It is obvious that bio-portal is the hub in life science cross domains. Cluster 1 and cluster 9 include specific semantic knowledge about bio-portal and gene ontology, while other clusters contain a relative comprehensive knowledge for each domain from different perspectives. Moreover, the visualization for each query has been implemented using CytoScape [45] in Figure 4.

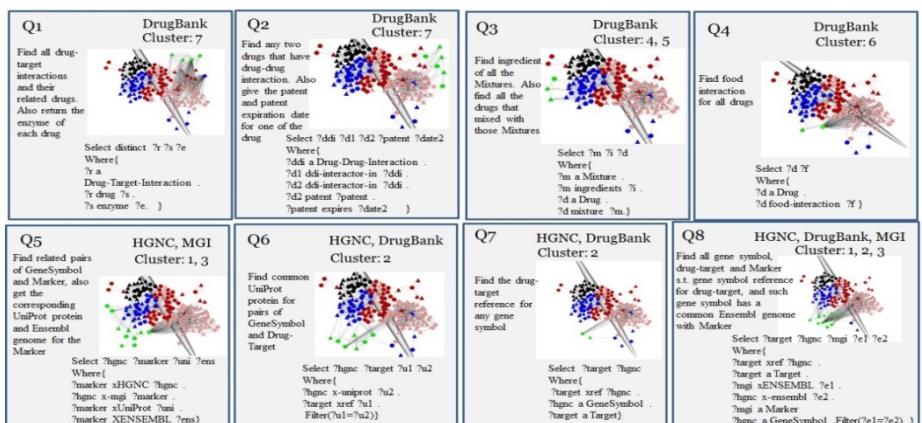

**Fig. 4.** Eight Queries with Query Graphs

## 8. Evaluation Plan

For future work, we plan to compare our proposed methods with previously developed tool BioBroker [46] with the fuzzy c-means clustering algorithm [47-49] on graph partition with semantic correlation. In addition, more biomedical heterogeneous ontologies and clinical narratives/literature will be incorporated to discover knowledge across biomedical and clinical domains [50-55]. Moreover, we will enable large graph analysis and process through adopting cloud computing techniques as well as distributed graph process [56].

For the future evaluation, there are three main lines of interests.

**Validation** The proposed approach will be validated by comparing with more existing algorithm and framework

**Scalability** Indexing approach will be applied to facilitate the management of big data. All datasts from Bio2RDF will be used to test the scalability of the proposed framework

**Benchmark** The existing Bio2RDF query sets will be analyzed to compare with our benchmark generation scheme in order to improve the coverage and semantic for each cluster association.

## 9. Reflections

Our approach overcomes the difficulties of less semantic across different knowledge domains by providing a new semantic distance measurement and clustering design. This way, we achieve grouping close domain knowledge together in order to help user design comprehensive query to fully discover cross-domain knowledge.